\documentclass[11pt]{article}

\usepackage{acl}

\usepackage{times}
\usepackage{tcolorbox}
\usepackage{pifont}

\usepackage[utf8]{inputenc} 
\usepackage[T1]{fontenc}   
\usepackage{hyperref}       
\usepackage{url}
\usepackage{multirow}       
\usepackage{booktabs}  
\usepackage{amssymb}
\usepackage{amsthm}

\usepackage{amsfonts}       
\usepackage{nicefrac}       
\usepackage{tabularx}
\usepackage{longtable}
\usepackage{supertabular}
\usepackage{makecell}
\usepackage{microtype}      
\usepackage{xcolor}         
\usepackage{tabularx}
\usepackage{graphicx}
\graphicspath{ {imgs/} }
\usepackage{svg}
\usepackage{enumitem}
\usepackage{subcaption}
\usepackage{amsthm, amsmath}
\usepackage[linesnumbered,ruled]{algorithm2e}
\usepackage{colortbl}

\definecolor{blue}{RGB}{17,220,247}
\definecolor{purple}{RGB}{163,115,250}
\definecolor{caribbeangreen}{rgb}{0.0, 0.8, 0.6}

\definecolor{GREEN}{RGB}{84,130,53}
\newcommand{\colorit}{\cellcolor{green!15}}

\newcommand{\colorg}{\cellcolor{gray!15}}

\usepackage{wrapfig}

\usepackage{todonotes}

\usepackage{bbm}

\usepackage{algorithm}
\usepackage{algpseudocode}

\newcommand{\name}{\textsc{Sunar}}

\definecolor{GREEN}{RGB}{84,130,53}

\usepackage{bbm}

\newcommand{\COT}{\textsc{Few-shot-cot}}
\newcommand{\self}{\textsc{self-ask}}

\newcommand{\zfshot}{\textsc{zero-shot-cot}}
\usepackage{pgfplots}
\pgfplotsset{compat=1.15}
\usepgfplotslibrary{
  statistics,
  colorbrewer,
  groupplots,
}
\usetikzlibrary{
  patterns,
  shapes.geometric,
  decorations.text,
  matrix,
  fit,
  backgrounds,
  positioning,
}
\tikzset{
  fignode/.style={
    outer sep=0.25em,
  }
}
\tikzset{
  framedfignode/.style={
    outer sep=0.25em,
    inner sep=0.5em,
    rounded corners,
    draw,
  }
}
\colorlet{plotColorNeutral}{gray}
\definecolor{plotColor1}{HTML}{f61a1c}
\definecolor{plotColor2}{HTML}{377eb8}
\definecolor{plotColor3}{HTML}{4daf4a}
\definecolor{plotColor4}{HTML}{984ea3}
\colorlet{plotColorNeutral*}{plotColorNeutral!40}
\colorlet{plotColor1*}{plotColor1!60}
\colorlet{plotColor2*}{plotColor2!60}
\colorlet{plotColor3*}{plotColor3!60}
\colorlet{plotColor4*}{plotColor4!60}
\pgfplotsset{
    colormap={greenred}{HTML=(4daf4a) HTML=(e41a1c)},
    colormap={redgreen}{HTML=(e41a1c) HTML=(4daf4a)}
}

\newcommand{\wqa}{\textsf{WQA}}

\newcommand{\bP}{\mathbb{P}}

\theoremstyle{definition}

\title{\name{}: Semantic Uncertainty based Neighborhood Aware Retrieval for Complex QA}

\author{ Venktesh V$^*$\\ TU Delft \\ \texttt{v.viswanathan-1} \\\texttt{@tudelft.nl}
\And Mandeep Rathee\Thanks{ Both authors contributed equally to this work} \\ L3S Research Center \\ \texttt{rathee} \\ \texttt{@l3s.de}  
\And  Avishek Anand \\ TU Delft \\ \texttt{avishek.anand}\\ \texttt{@tudelft.nl}  }

\begin{document}

\maketitle
\begin{abstract}

Complex question-answering (QA) systems face significant challenges in retrieving and reasoning over information that addresses multifaceted queries. While large language models (LLMs) have advanced the reasoning capabilities of these systems, the \textbf{bounded-recall} problem persists, where procuring all relevant documents in first-stage retrieval remains a challenge. Missing pertinent documents at this stage leads to performance degradation that cannot be remedied in later stages, especially given the limited context windows of LLMs which necessitate high recall at smaller retrieval depths.
In this paper, we introduce \name{}, a novel approach that leverages LLMs to guide a Neighborhood Aware Retrieval process.  \name{} iteratively explores a neighborhood graph of documents, dynamically promoting or penalizing documents based on uncertainty estimates from interim LLM-generated answer candidates. 
We validate our approach through extensive experiments on two complex QA datasets. Our results show that \name{} significantly outperforms existing retrieve-and-reason baselines, achieving up to a \textbf{31.84\%} improvement in performance over existing state-of-the-art methods for complex QA. 


\vspace{1.5em}
\includegraphics[width=1.25em,height=1.25em]{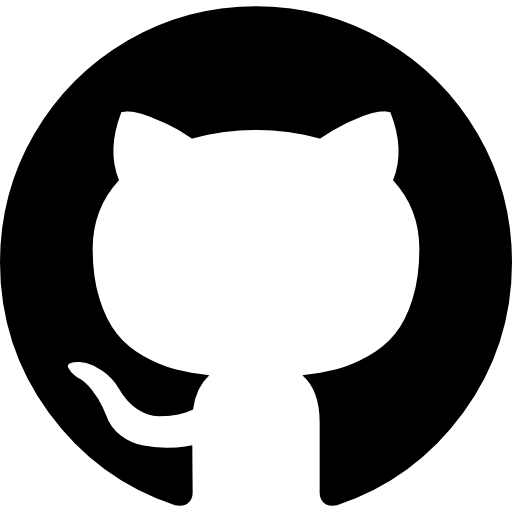}\hspace{.3em}
\parbox[c]{\columnwidth}
{
    \vspace{-.55em}
    \href{https://github.com/VenkteshV/SUNAR}{\nolinkurl{github.com/VenkteshV/SUNAR}}
}


\end{abstract}

\section{Introduction}
\label{chapter:introduction}

Open-domain complex question answering (CQA) has emerged as a critical challenge in natural language processing, demanding systems to comprehend, reason, and synthesize information over multiple queries and from multiple sources. 
While effective for simpler queries, traditional retrieve-and-reason pipelines struggle with the multifaceted nature of complex QA tasks. 
Complex QA systems typically involve multiple stages: query understanding, retrieval of relevant documents, and a reasoning phase that may or may not leverage large language models (LLMs). 
However, the effectiveness of these systems is typically bounded by the recall of the retrieval stage.
Specifically, if relevant documents are missed during this stage, they cannot be incorporated later, leading to a decline in overall system performance.

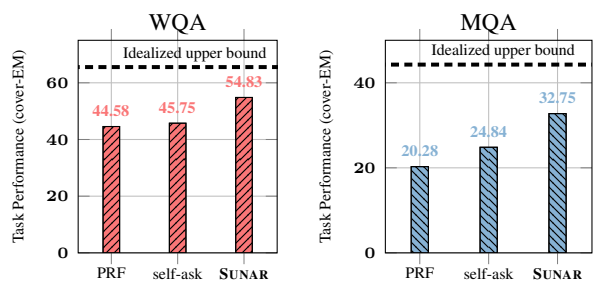
\begin{figure}[!t]
    \begin{subfigure}{.4\linewidth}
    \begin{tikzpicture}
\edef\mylst{"44.58","45.75","54.83"}
\edef\explora{"24.84","0","28.11","32.75","44.28"}

    \begin{axis}[
            ybar=5pt,
            width=4.2cm,
            height=4.4cm,
            bar width=0.25,
            every axis plot/.append style={fill},
            grid=major,
            xtick={1, 2, 3},
            xticklabels={ PRF, self-ask,\textbf{\name{}}},
            xlabel={},
            title={\small{WQA}},
            title style={yshift=-1.5ex}, 
            ylabel style = {font=\tiny},
        yticklabel style = {font=\boldmath \tiny,xshift=0.05ex},
        xticklabel style ={font=\tiny,yshift=0.5ex},
            ylabel={Task Performance (cover-EM)},
            enlarge x limits=0.25,
            ymin=0,
            ymax=75,
            nodes near coords style={font=\tiny,align=center,text width=2em},
            legend cell align={left},
            legend pos=north west,
            legend columns=-1,
    extra y tick style={grid=major,major grid style={thick,draw=black}}, 
            legend style={/tikz/every even column/.append style={column sep=0.5cm}},
        ]
        \addplot+[
            ybar,
            plotColor1*,
            nodes near coords=\pgfmathsetmacro{\mystring}{{\mylst}[\coordindex]}\textbf{\mystring},
            nodes near coords align={vertical},
            draw=black,
            postaction={
                    pattern=north east lines
                },
        ] plot coordinates {
                (1,44.58)
                (2,45.75)
                (3,54.83)
            };
            
        \draw[color=black, style=densely dashed,line width = 1.5pt] (axis cs:0,65.55) -- (axis cs:5,65.55)  node[pos=0.2, anchor=west, font=\tiny, yshift=5pt] {Idealized upper bound};           

    \end{axis}

\end{tikzpicture}
    \end{subfigure}
    \hspace{2em}
        \begin{subfigure}{.4\linewidth}
    \begin{tikzpicture}
\edef\mylst{"20.28","24.84","32.75"}

    \begin{axis}[
            ybar=5pt,
            width=4.3cm,
            height=4.4cm,
            bar width=0.25,
            every axis plot/.append style={fill},
            grid=major,
            xtick={1, 2, 3},
            xticklabels={ PRF, self-ask, \textbf{\name{}}},
            xlabel={},
            title={\small{MQA}},
            title style={yshift=-1.5ex}, 
            ylabel style = {font=\tiny},
        yticklabel style = {font=\boldmath \tiny,xshift=0.05ex},
        xticklabel style ={font=\tiny,yshift=0.5ex},
            ylabel={Task Performance (cover-EM)},
            enlarge x limits=0.25,
            ymin=0,
            ymax=50,
            nodes near coords style={font=\tiny,align=center,text width=2em},
            legend cell align={left},
            legend pos=north west,
            legend columns=-1,
    extra y tick style={grid=major,major grid style={thick,draw=black}}, 
            legend style={/tikz/every even column/.append style={column sep=0.5cm}},
        ]
        \addplot+[
            ybar,
            plotColor2*,
            draw=black,
            nodes near coords=\pgfmathsetmacro{\mystring}{{\mylst}[\coordindex]}\textbf{\mystring},
    nodes near coords align={vertical},
            postaction={
                    pattern=north west lines
                },
        ] plot coordinates {
                (1,20.28)
                (2,24.84)
                (3,32.75)
            };
            
        \draw[color=black, style=densely dashed,line width = 1.5pt] (axis cs:0,44.28) -- (axis cs:5,44.28)  node[pos=0.2, anchor=west, font=\tiny, yshift=5.0pt] {Idealized upper bound};

    \end{axis}

\end{tikzpicture}
    \end{subfigure}
 \vspace{-5pt}
\caption{The graph represents the comparison general decompose-retrieve paradigm for complex QA (\self{}), and the core approach of this work \name{} in open-domain setup for complex QA on 2WikiMultihopQA (WQA) (left) and MusiqueQA (MQA) (right).}
\label{fig:intro_plot}
\end{figure}

This \textit{bounded-recall problem} is more prominent in complex question answering where the initial question comprises multiple sub-questions that necessitate a retrieval for each sub-question~\cite{2wikimultihopqa,trivedi-etal-2022-musique}.
Failing to retrieve relevant documents to answer a sub-question, causes cascading failures that result in suboptimal performance of CQA systems.
Furthermore, the limited context windows of LLMs necessitate a focus on recall at smaller retrieval depths. 

\begin{figure*}
    \centering
    \includegraphics[width=0.7\linewidth]{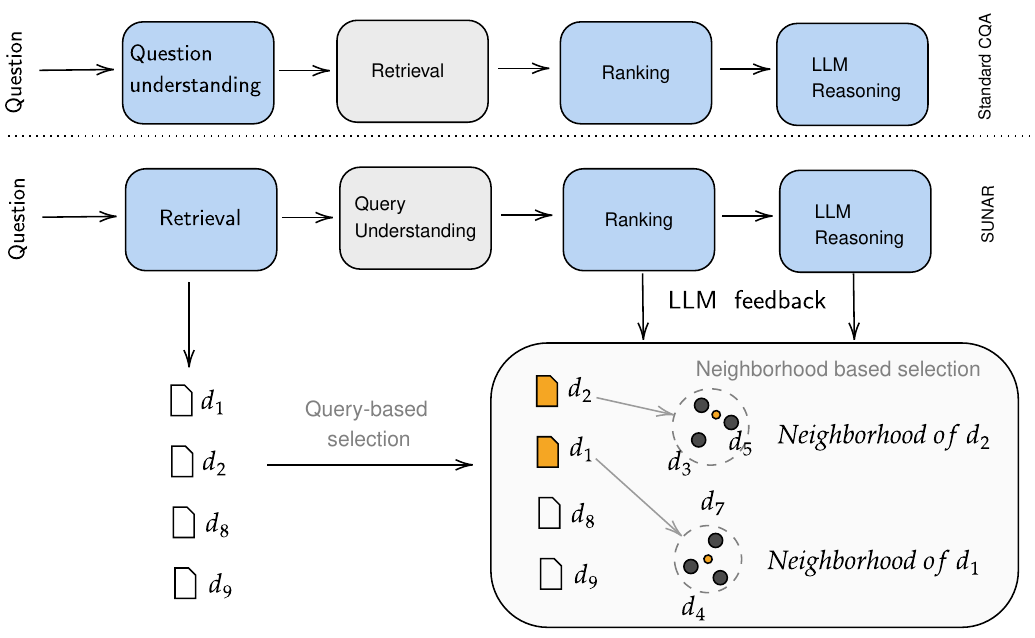}
    \caption{Overview of \name{} with Neighborhood Aware Retrieval (NAR) and LLM based feedback.}
    \label{fig:main}
\end{figure*}
Existing retrieval literature primarily focuses on improving precision at higher retrieval depths by either employing re-rankers~\cite{nogueira2019multistagedocumentrankingbert}, or using improving the LLM-reasoning by using self-correction mechanisms and decomposition strategies~\cite{self_ask,khattab2023dspycompilingdeclarativelanguage,asai2024selfrag}.
While these efforts led to notable improvements in performance in comparison to the standard open-domain QA baselines, the recall problem still persists as shown in Figure~\ref{fig:intro_plot}.

In this paper, we propose a novel method to bridge the recall gap by employing LLM feedback to guide the retrieval process. 
Our method is grounded on the clustering hypothesis, a well-known concept in information retrieval, that states that documents that are semantically similar to a relevant document tend to answer the same query~\cite{jardine1971use}.
Recently in the IR literature,  there have been initial attempts to exploit this idea to improve retrieval using better assessments from a re-ranker~\cite{macavaney2022adaptive,kulkarni2023lexically,rathee2024quam}.

We propose a neighborhood aware retrieval procedure that judiciously explores related documents of the initially retrieved documents leveraging the interim feedback of LLM-based answering system based on semantic uncertainty of potential answers.
In our approach, \name{} (\textbf{S}emantic \textbf{U}ncertainty based \textbf{N}eighborhood \textbf{A}ware \textbf{R}etrieval) we first compute a neighborhood graph based on semantic similarities between documents.
During the query processing phase, we judiciously explore the neighborhood graph of the retrieved documents to use in the LLM prompt as context.
Central to our method is the hypothesis that better document rankings correlate with lower uncertainty in the LLM-generated answer candidates. 
We quantify this uncertainty and use it as feedback to promote or penalize documents during retrieval dynamically. By integrating LLM-based signals into the neighborhood aware retrieval process, we effectively bootstrap the retrieval of relevant context, enhancing the overall system performance.

We conduct extensive experiments on two open-domain complex QA datasets to evaluate our approach. 
Our results show \textbf{significant improvements} over existing retrieve-and-reason baselines, with an increase in performance of up to \textbf{20\%} on WQA and up to \textbf{31.84\%} on MQA. 
To contextualize our performance, we close the performance gap with an idealized upper bound from $43\%$ to $19\%$ on WQA~\footnote{An idealized upper bound is a QA system which has perfectly relevant results as context}.
Our experiments also show that \name{} can be retrofitted to LLM-based reasoning approaches~\cite{yao2023react,xu2024searchinthechaininteractivelyenhancinglarge}, to improve their performance further.








\section{Related Work}
\subsection{Complex Question Answering}
Complex Question Answering tasks require multi-step reasoning by leveraging information from multiple sources \cite{2wikimultihopqa,trivedi-etal-2022-musique,multi_hop_survey}. HotPotQA \cite{yang-etal-2018-hotpotqa}, one of the first datasets introduced for multi-hop complex QA, which however has been found to not necessitate multi-hop reasoning \cite{min-etal-2019-compositional,trivedi-etal-2022-musique}. To tackle this, more challenging datasets that require connected/multi-hop reasoning like 2WikiMultiHopQA \cite{2wikimultihopqa} and MusiqueQA \cite{trivedi-etal-2022-musique} have been proposed.
Supervised and unsupervised~\cite{perez-etal-2020-unsupervised}  approaches for complex QA often rely on fine-tuning multiple specialized models to iteratively refine the query representations and arrive at final answer~\cite{perez-etal-2020-unsupervised,multi_hop_survey}. However, with the recent advances in Large Language Models (LLMs) \cite{wei2022emergent,brown2020gpt3,chainofthought}, they are being adopted as general-purpose answer engines for complex tasks, obviating the need for fine-tuning of multiple models.
However, relying only on parametric knowledge of LLMs may lead to sub-par performance due to factual inconsistencies resulting from hallucination \cite{hallucination}.
\vspace{-0.7em}
\subsection{Retrieval Approaches}

The typical retrieval approaches for QA systems follow a retrieve-and-read framework~\cite{chen-etal-2017-reading}, where the retriever can be either sparse~\cite{BM25,splade} or dense~\cite{karpukhin-etal-2020-dense,ance,mpnet}.
The reader or the answering system can be either fine-tuned~\cite{multi_hop_survey} or employ few-shot or zero-shot reasoning~\cite{brown2020gpt3,lewis2021retrievalaugmented}.
Pseudo Relevance Feedback (PRF) approaches aim to bridge query-document vocabulary/representation mismatch and improve retrieval effectiveness by recomputing query representations for re-retrieval/re-ranking using relevant documents from first-stage retrieval \cite{ance_prf}. 
However, PRF methods are known to suffer from \textit{drift} resulting in suboptimal performance due to retrieving irrelevant contexts.
Our proposed method combines standard retrieve and rerank systems with documents neighborhood information. Most similar to our work, adaptive retrieval approaches have been proposed \cite{macavaney2022adaptive,kulkarni2023lexically, rathee2024quam} that use rankers to guide the retrieval process while keeping the re-ranking costs constant. 
We are different from earlier work in adapting LLM feedback for neighborhood retrieval. Also different from PRF techniques, we operate on the document similarity dimension instead of similarity to the query representation.


\subsection{Pipelines for complex QA}

With the recent advances in LLMs, off-the-shelf Retrieval Augmented Generation (RAG) approaches have been proposed which retrieve and incorporate external knowledge to mitigate factual inconsistencies and generate accurate answers \cite{khot2023decomposed,xu2024searchinthechaininteractivelyenhancinglarge,schick2023toolformerlanguagemodelsteach}. These RAG pipelines are coupled with query understanding and reasoning approaches that leverage the emergent capabilities of the LLM to decompose a complex question into sub-questions and solve them iteratively \cite{self_ask,xu2024searchinthechaininteractivelyenhancinglarge,khot2023decomposed,yao2023react,trivedi-etal-2023-interleaving,dua-etal-2022-successive}.  However, the existing RAG pipelines for complex QA employ off-the-shelf retrievers for retrieving relevant knowledge without comprehensive evaluation and are limited by the performance of the retriever \cite{asai2024selfrag,li-etal-2024-self-prompting}. An additional challenge is the context limit of LLMs employed as readers, making it necessary to capture the relevant documents within the limited budget.  Additionally, the inherent noise in retrieved documents, could result in a wrong reasoning path and \cite{shi-etal-2024-generate} and derive a wrong answer. Hence, in this work, we focus on improving the recall of top-ranked documents that fit into the context limit of the reader LLMs through document Neighborhood Aware Retrieval enhanced with semantic uncertainty based LLM feedback. 
\section{Methodology}
\label{sec:methods}

The common approach to performing open-domain QA tasks typically follows a two-step process: retrieval of relevant contexts, followed by a reasoning or QA step using those retrieved contexts. 
This is exemplified by traditional RAG pipelines~\cite{lewis2021retrievalaugmented}.
For simpler or direct questions, this method is often sufficient. However, complex questions tend to consist of multiple sub-questions, which requires more advanced query understanding steps, such as decomposing the main question into sub-questions prior to retrieval.
In this context, given a complex question $q \in Q$ and its corresponding answer $a \in A$, the decomposition process can be formally defined as:
$de(q, a) = \{(sq_1, R_1, sa_1), \dots, (sq_k, R_k, sa_k)\}
$
where $k = k_{(q, a)}$ represents the number of sub-questions, $(sq_i, sa_i)$ is the pair of sub-question and corresponding answer, and $R_i$ is a list of top-$l$ ranked documents (with $l \leq 10$) supporting the sub-question $sq_i$.
After decomposing the complex question into sub-questions, the retrieval step is followed by a ranking phase, typically performed by a fine-tuned model. This ranking is critical due to the limited input budget of large language models (LLMs), which restricts the amount of information they can process at once. The ranking helps prioritize the most relevant documents for the final reasoning step.
Finally, in the reasoning phase, the LLM processes the ranked and retrieved contexts to generate the final answer. In this work, we utilize the LLM as the core question-answering model, responsible for reasoning over the retrieved and ranked contexts and providing the final answer. An overview of our approach is shown in Figure \ref{fig:main}. In our approach, we employ a Neighborhood Aware Retrieval (NAR) method to overcome the bounded-recall issue of first-stage retrieval in classic RAG pipelines. To further ensure we capture the most relevant documents in the limited top-$l$ budget we employ uncertainty-based LLM feedback to rescore the documents.

\algdef{SE}[DOWHILE]{Do}{doWhile}{\algorithmicdo}[1]{\algorithmicwhile\ #1}%
\begin{figure}[t]\vspace{-1em}
\begin{algorithm}[H]
  \SetAlgoLined
\caption{The \name{} Algorithm} 
\label{alg: main}
\begin{algorithmic}[1]
\small
\Require Initial retrieved list $R$, batch size $b$, re-ranking budget $c$, document graph $G$
\Ensure Re-Ranked pool $R^+$
\State $R^+ \gets \emptyset$ \Comment{Re-Ranking results}
\State $C \gets R$ \Comment{Candidate pool}
\State $N \gets \emptyset$ \Comment{Neighbor pool}
\Do
  \State $B \gets$ \Call{Score}{top $b$ from $C$, subject to $c$} 
  
    \State  $\{sa_1...sa_m\} \gets \phi(\bP_{LLM}(sq_1,B))$   

    \State  $\{ac_1..ac_s\} \gets \sigma(sa_1..sa_m)$ 
    \Comment{Clustering}

    \State
 \State $B \gets$ \Call{rescore}{$B$,$1/s$} \Comment{Rescore batch}
  \State $R^+ \gets R^+ \cup B$ \Comment{Add batch to results}

\State
  \State 
  // Discard Batches
  \State $R \gets R \setminus B$ 
  \State $N \gets N \setminus B$ 
  \State $N \gets N \cup (\Call{Neighbours}{B, G} \setminus R^+$) 

\State
\State //Alternate $R$ and $N$
  \State $C \gets \begin{cases} 
      R & \text{if}\; C = F   \\
      N  & \text{if}\; C = N \\
   \end{cases}$ 
\doWhile{$|R^+| < c$}
\end{algorithmic}
\end{algorithm}
\end{figure}



\subsection{Neighborhood Aware Retrieval}
\label{sec:adaptive_retrieval}

Current RAG pipelines usually employ an off-the-shelf first-stage retriever for fetching relevant knowledge. 
While query understanding /question decomposition can help fetch better documents, the performance of complex QA is still limited by the quality of the first-stage retriever. 
To overcome the recall limitations of first-stage retrieval, we adopt Neighborhood Aware Retrieval (NAR) which employs the \textit{Clustering Hypothesis}~\cite{jardine1971use} which suggests that documents in the vicinity of highly scored documents tend to answer the same queries/questions.

\subsubsection{Neighborhood Graph Construction} 

NAR first constructs a document-neighborhood graph $G = (V, E)$, also called the neighborhood graph in an offline phase. 
Each document is a node and the top-$k$ nearest neighbors are the edges, documents which are in the close vicinity (hence, $|E|=k|V|$) according to semantic relatedness. The neighborhood graph is constructed using a dense or sparse retriever. In the case of a dense retriever, for example, ColbertV2, the neighborhood graph can be constructed by using the representation space generated by the document encoder and fetching $k$-nearest neighbor documents for each document in the corpus. 
The neighborhood graph is constructed using a dense retriever if sparse retrieval is used for first-stage retrieval and vice versa to capture complementary signals. Note that the neighborhood graph construction is a one-time activity for each document collection, and during inference, the indexed graph is directly used for lookup for efficiency. 

\subsubsection{The SUNAR Algorithm} NAR starts with an initial set of ranked documents $R$ obtained from first-stage retrieval. NAR scores the documents in batches, each of size $b$, in each iteration till budget $c$ is reached (stopping condition in Line 15 of Algorithm \ref{alg: main}. NAR comprises a dynamically updated candidate documents pool $C$ (initialized to $R$, line 2) and a dynamically adapted neighbor pool ($N$, initially empty, line 3). Firstly, NAR employs a cross-encoder based re-ranker to score top $b$ documents from $C$ which constitute a batch $B$ (line 5). After this step, in LLM guided NAR (lines 6-9) we employ LLM-based feedback to re-score the documents in batch $B$. While we further elaborate on this in Section \ref{sec:answer_uncertainty}, vanilla NAR does not include this re-scoring step and adds $B$ to $R^+$ (line 9). Then the documents of the batch $B$ are removed from $N$ and $R$ as they are already ranked. Following this step, the neighbors of documents in $B$ are looked up from the neighborhood graph $G$. These documents, barring those already ranked are added to the neighbor pool (line 13) prioritized according to the ranking score of the source document in $B$. NAR then explores a batch of documents from neighbor pool $N$ instead of the next batch from $R$. This ensures the final ranked list includes documents not only from $R$ but additional documents also that are not included in the initial retrieval $R$. This aids in overcoming the recall limitations of first-stage retrieval. NAR proceeds by alternating between $R$ and $N$ till budget $c$ is reached and the final ranked list $R^+$ is returned. Note that NAR is invoked for each sub-question $sq_i$. 
\subsection{LLM feedback for Neighborhood Aware Retrieval}
\label{sec:answer_uncertainty}
To improve the re-ranking of documents and promote the most relevant documents in the limited top-$l$ budget (for instance, where $l$=10), 
we propose an LLM-based feedback mechanism to re-score the top-ranked documents from the cross-encoder for each batch of NAR. While the cross-encoder aids in re-rank the retrieved documents for each sub-question from a relevance perspective, the LLM feedback helps re-score documents from an uncertainty perspective.

More formally, given the sub-question and ranked list of documents $B$ from the batch in the current iteration of NAR  $(sq_1, B)$, where $B=[d_1,d_2,...,d_b]$, the ASU (Answer Semantic Uncertainty) re-scores the returns the batch $B$ with updated scores (6-9 of Algorithm \ref{alg: main}).
\begin{equation}
B = \mathtt{LLM\_SCORE}(sq_1,B)
\label{eq:rescore}
\end{equation}

It accomplishes this by first estimating the consistency of the answers based on the number of answer semantic sets. This is computed by first estimating multiple answers for a given sub-question and ranked document list
\begin{equation}
   \{sa_1..sa_m\} = \phi(\bP_{LLM}(sq_1,R),m)
   \label{eq:response_gen}
\end{equation}
where $\phi$ is the response generator that generates multiple responses ($sa_1...sa_{m}$) from the distribution $\bP_{LLM}$. We further define $\sigma$ as the estimator of answer semantic sets through semantic equivalence estimation for semantic clustering. A semantic set is defined as a set of sequences that share the same meaning following the work \cite{kuhn2023semantic}. $\sigma$ takes as input a set of answers and outputs the answer semantic sets. 

\begin{equation}
\{ac_1..ac_s\} = \sigma(\{(sa_1..,sa_m)\})
\label{eq:sigma}
\end{equation}
where $s$ denotes the number of semantic sets

It accomplishes this by clustering equivalent answers to the same answer semantic set through bi-directional entailment computed between each pair of answers. We employ bi-directional entailment as it is a stronger criterion for semantic equivalence \cite{kuhn2023semantic}. $\sigma(sa_i, sa_j)$ is defined as:
\[
   M_{NLI}(sa_i \rightarrow sa_j) \land M_{NLI}(sa_j \rightarrow sa_i)
\]

 The entailment condition is evaluated for each pair of sequences. If both directions hold true, \( sa_i \) and \( sa_j \) are assigned to the same semantic set. If the entailment fails in either direction, the sequences are placed in distinct semantic sets.  We posit that more semantic sets indicate that the LLM is uncertain about the answer to the given question and the current batch of documents. Finally, we re-score the documents in the current batch by penalizing the scores from the cross-encoder as follows: 
 
 $ \mathtt{LLM\_SCORE}(sq_1,B) = [\frac{sc_1}{s}.. \frac{sc_n}{s}]
      \label{eq:re-score-2} $ where $sc_i$ refers to the cross-encoder score for the ith document for a given sub-question in batch $B$. The above process is repeated for all batches till the budget $c$ for NAR is reached for a given sub-question. This is followed by the answer generation step using the LLM by leveraging top-$l$ (where $l$=10) documents from the re-ranked list. The same is repeated for all sub-questions till the decomposition stops and the final answer is generated.

 However, we observe that in this sequential reasoning process, the final answer is derived solely based on answers to previous sub-questions and only the evidence for the last sub-question. Hence, any errors in intermediate steps could result in cascading errors resulting in a wrong final answer. Inspired by post-hoc LLM correction strategies, we propose a simple yet significantly effective Meta Evidence Reasoner (MER) to tackle this issue. The MER component leverages the reasoning path obtained through sequential reasoning $[(sq_1,sa_1...(sq_n,sa_n)]$ and the top-$l$ evidences from the set of ranked list of documents ${R_1^+,R_2^+...}$ across sub-questions and prompts the LLM with the original question to obtain the final answer. The prompt is as shown in Figure \ref{fig:meta} in Appendix \ref{app:prompts}.  

\section{Experimental Setup}
 \label{sec:experiments}

 We answer the following research questions:
 
 \noindent\textbf{RQ I.} Can NAR improve the recall for better complex QA?
 
\noindent\textbf{RQ II.} Can LLM-guided NAR improve overall complex QA performance compared to existing state-of-the-art?

\noindent\textbf{RQ III.} How well does \name{} work across different LLM substrates and with different reasoning approaches?

\textbf{Datasets:} We experiment on well-known QA datasets that require compositional reasoning, namely MuSiQue (MQA) \cite{trivedi-etal-2022-musique} and 2WikiMultiHopQa (WQA) \cite{2wikimultihopqa}. It has been observed that multi-hop datasets like HotpotQA \cite{yang-etal-2018-hotpotqa} do not necessitate connected reasoning and can be answered using single hop \cite{min-etal-2019-compositional,trivedi-etal-2022-musique}. However, MQA and WQA are designed in a manner that mandates compositional/multi-step reasoning. Following \cite{self_ask}, we employ the 2-hop questions from MQA leading to 1252 questions for evaluation. For WQA we evaluate on the sampled 1200 questions following \self{} \cite{self_ask}. Combining the contexts with distractors from MQA and WQA following the setup of \cite{khot2023decomposed} results in a corpus of 569461 documents. 

\noindent \textbf{Evaluation Metrics:} We evaluate the QA performance using the well-known cover-EM metric \cite{cover_em,self_ask} which checks whether the ground truth is contained in the generated answer. We also evaluate retrieval performance for the original questions as shown in Tables \ref{table:comparison_retrieval},\ref{table:retrieval_performance} using Normalized Discounted Cumulative Gain (nDCG@k) and Recall@k (R@k). 


\begin{table}[!t]
    \centering
    \small
    \setlength{\tabcolsep}{2.0pt}
    \begin{tabular}{lccc}
    \toprule
     \textbf{Method}& \multicolumn{1}{c}{MQA}& \multicolumn{1}{c}{\wqa{}}\\

     \midrule
          \colorg \textbf{Methods (w/o query understanding)} & \colorg& \colorg& \colorg\\
     
     \zfshot{} \cite{kojima2023large}  & 8.62 & 30.42 \\
     \COT{} \cite{wei2023chainofthought} &15.02& 32.83 \\

               \COT{} +PRF \cite{ance_prf} & 16.69 & 35.55 \\
$\name{}_R$ (ours) & 21.32 & 40.96 \\

    \hline
              \colorg \textbf{Methods (w/ query understanding)} & \colorg& \colorg& \colorg\\


    Self-RAG \cite{asai2024selfrag} & 17.80 & 35.25 \\
    ReAct \cite{yao2023react} & 21.41 & 43.25  \\

    DecomP \cite{khot2023decomposed} & 21.01 & 44.08 \\
            IRCoT \cite{trivedi-etal-2023-interleaving} & 24.12 & 42.01 \\
        SearChain \cite{xu2024searchinthechaininteractivelyenhancinglarge} & 21.72& 44.42  \\ 
            \self{} +PRF \cite{ance_prf} & 20.28 & 44.58 \\

            \self{} \cite{self_ask} & 24.84& 45.75 \\



        \hline
                 \colorg \textbf{NAR (w/ query understanding) (ours)}& \colorg& \colorg& \colorg\\

        $\name{}_{R}$ & 28.11 & 47.67\\  
        \name{} & \textbf{32.75} $\dagger$& \textbf{54.83}$\dagger$ \\
        \hline  \hline
                 \colorg \textbf{Golden Evidence }& \colorg& \colorg& \colorg\\   
                                  \colorg \textbf{(Ideal Upper Bound)}& \colorg& \colorg& \colorg\\
                 \COT{} & \colorit 44.28     &  \colorit  65.55  \\

     \bottomrule
    \end{tabular}
    \caption{Results across datasets. The model used for \name{} and other approaches is gpt-3.5-turbo unless otherwise specified.$\dagger$ indicates statistical significance over \self{} at 0.1 level.}
    \vspace{-2pt}
    \label{tab:main_result}
\end{table}

\noindent \textbf{\name{} details}: We employ SPLADE-v2 for the first stage retrieval in \name{} as it is better than other off-the-shelf dense retrievers usually employed in RAG pipelines as seen from results in Table \ref{table:retrieval_performance} in \textbf{Appendix \ref{app:retrieval_perf}}. The neighborhood graph for \name{} is formed using ColbertV2 which is a late-interaction model and is better at capturing relevance between documents which are usually longer than queries. We perform a KNN search over the corpus for each document and retrieve the top-100 documents as neighbors to form the neighborhood graph. In the NAR setup, we consider top-100 documents from the first stage retrieval and only 10 neighbors for the documents under consideration in the current iteration for a fair comparison with baselines that re-rank top-1000 documents from first stage retrieval (Table \ref{tab:ranking} in Appendix \ref{app:ranking} and PRF baselines in Table \ref{tab:main_result}). We use \textit{nreimers/mmarco-mMiniLMv2-L12-H384-v1} as scorer in NAR pipeline and for re-ranking enhanced baselines. In the end, the top-10 documents are used from the ranked list for LLM reasoning to answer the question in \name{} and all baselines. To observe the impact of variation on the number of documents we experiment with \{1,3,5,7,10\} documents for WQA and report the performance in Figure \ref{fig:top_k} in Appendix \ref{app:top_k}. We observe top-10 to be optimal from the results.  While, \name{} uses \self{} style decomposition for query understanding, we also show it generalizes to other approaches like SearChain and ReAct (Table \ref{tab:reasoning_substrates}). We also evaluate variations of \name{} where, $\name{}_R$ refers to w/o ASU and MER components.

\noindent \textbf{LLM Details:} We employ gpt-3.5-turbo as the backbone for the baselines and our approach for results shown in Table \ref{tab:main_result}. We also evaluate the best baseline and our approach using alternate models like gpt-4o-mini and open-source models like LLama 3.1 and Mistral v0.2.
We set max\_tokens for output generation to 1000 and $frequency\_penalty$ to 0.8 and $presence\_penalty$ to 0.6 to reduce repetition. 
 \textbf{Baselines:}
We compare \name{} with state-of-the-art pipelines that employ retrieval without query understanding (w/o indicates without) and approaches that employ question understanding. \COT{} \cite{wei2023chainofthought} employs stepwise reasoning given the original question and evidence retrieved using SPLADEv2 for the original question. We also compare with state-of-the-art query understanding or reasoning approaches like \name{} \cite{self_ask}, DecomP \cite{khot2023decomposed}, SearChain \cite{xu2024searchinthechaininteractivelyenhancinglarge} that decompose the original question and interactively query the retriever to obtain relevant contexts. We use SPLADEv2 for first-stage retrieval across all baselines for a fair comparison. While all baselines use retrieve-reason paradigm, we further enhance these baselines by re-ranking the contexts and results are reported in Appendix \ref{app:ranking} (Table \ref{tab:ranking}). We further elaborate on PRF baseline in Appendix \ref{app:baselines}.
All prompts are presented in Appendix \ref{app:prompts}.

\section{Results}
\label{sec:results}
\subsection{Impact of Neighborhood aware retrieval}
To answer \textbf{RQ1}, we evaluate the impact of NAR (neighborhood Aware Retrieval) to state-of-the-art approaches adopted for complex reasoning, as shown in Table \ref{tab:main_result}. We observe that $\name{}_{R}$ which employs NAR outperforms all existing approaches, highlighting the ability of NAR to surface relevant documents in the top-$l$ results where $l$ <= 10. We observe that $\name{}_{R}$ significantly outperforms \COT{}+PRF  in both w query understanding and w/o query understanding setups. This is primarily because in NAR document-document similarities are also captured, rather than depending only on query-document similarity. Hence, a document that is less related to the current query might be closer to another relevant document not currently captured in the top-$l$ results. NAR helps capture such documents and re-score them to surface them in the top-$l$ results.  We posit that \COT{}+PRF underperforms due to the possibility of drift where the augmented query may capture only relevant documents for a particular aspect missing out on other relevant documents due to the multi-aspect nature of complex QA. We also observe that $\name{}_{R}$ outperforms other state-of-the-art approaches that perform decomposition. These approaches like \self{}, ReAct and Searchain decompose the query into sub-questions that capture the multi-aspect nature of the query and perform better than approaches without decomposition. However, they still rely on query-document relevance estimates from first-stage retrieval and do not optimize for recall at top-$l$ and hence fall short when compared to $\name{}_{R}$ which captures more relevant documents with the limited budget. Even when these query understanding baselines are enhanced with re-ranking, they fall short compared to \name{} as observed from Table \ref{tab:ranking} due to the bounded-recall problem.
\subsection{Significance of LLM guided NAR for complex QA}

\begin{table}[!t]
    \small
    \centering

    \begin{tabular}{lccc}
    \toprule
     \textbf{Method}& \multicolumn{1}{c}{MQA}& \multicolumn{1}{c}{\wqa{}}\\

     \midrule
          \colorg \textbf{w/o \name{}} & \colorg& \colorg& \colorg\\

        ReAct &  21.41 & 43.25  \\
SearChain & 21.72 & 44.42 \\
              \colorg \textbf{w/ \name{}} & \colorg& \colorg& \colorg\\

        ReAct & 28.67 & 48.40 \\

      SearChain & \textbf{30.43} & \textbf{55.50} \\
     
     \bottomrule
    \end{tabular}
    \caption{Performance comparison when augmenting \name{} to other approaches (gpt-3.5-turbo).}
    \vspace{-2pt}
    \label{tab:reasoning_substrates}
\end{table}

\begin{table}[!t]
    \centering
    \small
    \begin{tabular}{lccc}
    \toprule
     \textbf{Method}& \multicolumn{1}{c}{MQA}& \multicolumn{1}{c}{\wqa{}}\\

     \midrule
          \colorg \textbf{gpt-4o-mini} & \colorg& \colorg& \colorg\\

        \self{} & 26.76  & 37.33 \\
     \name{} & \textbf{32.19} & \textbf{48.16} \\

              \colorg \textbf{Llama 3.1 (8B)} & \colorg& \colorg& \colorg\\

        \self{} & 5.43& 25.83 \\
     \name{} & \textbf{13.82} & \textbf{39.52} \\

                   \colorg \textbf{Mistral v0.2 (7B)} & \colorg& \colorg& \colorg\\

        \self{} & 7.84 & 27.72\\
     \name{} & \textbf{26.12}& \textbf{40.23}\\
     
     \bottomrule
    \end{tabular}
    \caption{Results across different LLM substrates.}
    \label{tab:llm_substrates}
\end{table}

\begin{table}[!t]
    \centering
    \small
    \begin{tabular}{lccc}
    \toprule
     \textbf{Method}& \multicolumn{1}{c}{MQA}& \multicolumn{1}{c}{\wqa{}}\\

     \midrule
     \name{} & \textbf{32.75} & \textbf{54.83} \\
    \name{} (-ASU) & 30.43& 50.75 \\

        $\name{}_{R}$ (-ASU) (-MER) & 28.11 & 47.67 \\




     
     \bottomrule
    \end{tabular}
    \caption{Ablations of \name{} (gpt-3.5-turbo).}
    \label{tab:ablations}
\end{table}

\begin{table}[h!]
\small
\centering
\setlength{\tabcolsep}{3.0pt}
\begin{tabular}{p{2.8cm}cccc}
\toprule
\multirow{2}{*}{\textbf{Retriever}}  & \multicolumn{2}{c}{\textbf{MQA}} & \multicolumn{2}{c}{\textbf{WQA}}  \\
\cmidrule{2-5}
 & R@1 & R@10 &R@1 & R@10 \\
\midrule
\multirow{1}{*}{\self(Re-Rank)} &  0.157 & 0.309  & 0.230   & 0.402   \\
\self{}(Retrieval) & 0.152 & 0.240& 0.205& 0.399 \\
$ \name{}$ & \textbf{0.284}& \textbf{0.459}& \textbf{0.287} & \textbf{0.606} \\

\bottomrule
\end{tabular}
\caption{Retrieval performances on MQA and WQA  with query understanding. }
\label{table:decomposed_comparison_retrieval}
\end{table}

\begin{table*}[hbt!]
\small
    \setlength{\tabcolsep}{1.5pt}
    \begin{tabular}{p{2cm}p{14cm}}
\toprule
    \textbf{Method}     & \textbf{Evidences} \\
\midrule

\midrule

\colorg \textbf{Question}  & \colorg \textbf{Where was the director of film Ronnie Rocket born?} [Dataset: \textbf{WQA}] \\
\self{}  & [Evidence 1]:  \textcolor{red}{This is a list of film series by director.}  \\
& [Evidence 2]:  \textcolor{red}{This is a list of notable directors in motion picture and television arts.} \\
& [Final Answer]: \textcolor{red}{Unknown} \\
 \name{} (ours)  & [Evidence 1]:  \textcolor{teal}{Ronnie Rocket is an unfinished film project written by David Lynch, who also intended [\dots].}  \\
& [Evidence 2]:  \textcolor{teal}{David Keith Lynch was born in Missoula, Montana, on January 20, 1946. His father [\dots] .} \\
& [Final Answer]: \textcolor{teal}{Missoula, Montana}
 \\

\midrule

\colorg \textbf{Question}  & \colorg \textbf{Who did the screenwriter for Good Will Hunting play in Dazed and Confused?} [Dataset: \textbf{MQA}] \\
\self{}  & [Evidence 1]:  \textcolor{red}{Damon begins working alongside his younger brother, Stefan Salvatore, to resist greater[\dots].}  \\
& [Evidence 2]:  \textcolor{red}{Damon Salvatore is a fictional character in The Vampire Diaries. He is portrayed by Ian Somerhalder in the television.} \\
& [Final Answer]: \textcolor{red}{Damon Salvatore} \\
 \name{} (ours)  & [Evidence 1]:  \textcolor{teal}{Damon and Ben Affleck wrote \"Good Will Hunting\" (1997), a screenplay[\dots].}  \\
& [Evidence 2]:  \textcolor{teal}{Benjamin Affleck- Boldt( born August 15, 1972) is an American actor . He later appeared in the independent coming- of- age comedy\" Dazed and Confused\ as Fred O'Bannion [\dots]"} \\
& [Final Answer]: \textcolor{teal}{Fred O'Bannion}
 \\
\midrule

\end{tabular}
\caption{Qualitative analysis of evidences  retrieved by \self{} and \name{}}
\label{tab:qualitative}
\end{table*}

To address \textbf{RQ2}, we evaluate \name{} to observe the impact of LLM guidance on NAR and downstream LLM reasoning for complex QA. As observed from Table \ref{tab:main_result}, \name{} significantly outperforms state-of-the-art approaches for complex QA and also outperforms $\name{}_{R}$ by about \textbf{15.02\%} in WQA and by \textbf{16.9\%} in MQA. We observe that this is primarily due to the LLM guidance in the form of Answer Semantic Uncertainty (ASU) based re-scoring of documents. This is also evident from retrieval performance reported in Table \ref{table:decomposed_comparison_retrieval}, where \name{} outperforms \self{} by \textbf{48.54 \%} on MQA and by \textbf{50.75\%} on WQA which also positively impacts downstream reasoning performance. While NAR ($\name{}_{R}$) does offer improvements over traditional RAG pipelines for complex QA, it still relies on the notion of relevance from the cross-encoder employed as a re-ranker. \name{} employs LLM based ASU measure as feedback that goes beyond semantic relatedness and penalizes documents leading to inconsistent answers (more semantic sets). This helps reduce the number of distractors in top-$l$ results and ASU also serves as a proxy of whether the evidences help provide consistent answers for the question. Other query understanding approaches like SearChain, ReAct, and \self{} decompose the question and iteratively interact with the retriever to obtain relevant documents but do not handle the presence of distractors as they purely rely on semantic relatedness to the query.  For instance, in our \textbf{qualitative analysis} in Table \ref{tab:qualitative}, we observe that on MQA \self{} recovers documents irrelevant to answer the question but which have high semantic relatedness to the sub-question related to ``Matt Damon". This results in wrong answer as they are distractors and do not answer the complex question. However, \name{} is able to surface two of the most relevant documents that directly answer the question in the top-5 results. Hence, \name{} outperforms other approaches owing to LLM guided feedback for retrieval through ASU based penalty and LLM guided feedback for reasoning through MER based post-hoc correction.

\vspace{-0.6em}
\subsection{ \name{} with different LLM substrates and query understanding approaches}
To answer \textbf{RQ3}, we compare \name{} with the best baseline \self{} on different LLM substrates such as gpt-4o-mini, Llama 3.1 (8B), and Mistral v0.2 (7B) to observe the generalization capabilities of \name{}. From the results in Table \ref{tab:llm_substrates}, we observe that \name{} performs significantly better, demonstrating that \name{} is LLM agnostic and offers significant gains over existing state-of-the-art query understanding and RAG pipelines on smaller open-source models. We observe that this is primarily due to the relevant evidence surfaced by LLM guided NAR with fewer distractors.

 To demonstrate that \name{} is agnostic to query understanding/reasoning approaches, we combine \name{} with approaches like ReAct and SearChain. The results are as shown in Table \ref{tab:reasoning_substrates}. We observe that ReAct and SearChain augmented with \name{} offer significant gains over their counterparts that use existing retrieval approaches. 

\vspace{-0.7em}
\subsection{Ablations of \name{}}
We perform several ablations to assess the impact of LLM feedback for retrieval (ASU) and LLM feedback for reasoning (MER). The results are as shown in Table \ref{tab:ablations}. We observe that both ASU based penalty (LLM feedback for retrieval) and MER based post-hoc correction (LLM feedback for reasoning) are essential for superior performance on MQA and WQA as \name{} outperforms the variant without ASU and $\name{}_{R}$. We observe that removing LLM feedback for retrieval results in more distractors and hence a drop in performance. Further, $\name_{R}$ which employs only NAR without any form of LLM feedback has a significant drop in performance due to the presence of distractors in ranked documents from NAR and cascading errors in the stepwise reasoning due to lack of post-hoc correction using MER.

\section{Conclusion}

In this paper, we introduce a novel approach \name{} to improve retrieval and reasoning for complex QA. We propose Neighborhood Aware Retrieval (NAR) to improve recall in retrieved documents for downstream reasoning. We further augment NAR with LLM based feedback to reduce the impact of distractors and improve downstream reasoning. We observe significant performance improvements over state-of-the-art approaches on different datasets across different LLMs. Furthermore, when \name{} is applied to other reasoning/query understanding approaches it leads to significant performance improvements demonstrating that \name{} is agnostic of query understanding approaches and LLMs. In the future, we plan to explore further efficient methods for the ASU component.  
\section{Limitations}
Although \name{} offers significant performance improvements over existing off-the-shelf RAG pipelines, there are a few limitations that offer avenues for future work. Though NAR is based on the philosophy of adaptive retrieval which is more efficient than exhaustive re-ranking, a more principled selection of candidates for ranking in line 18 of Algorithm \ref{alg: main} with early stopping criterion could make it more efficient. We defer this for future work as it is beyond the scope of current work. For instance, the LLM based feedback now relies on computing semantic sets, which serve as a proxy for consistency in LLM answers for a given set of evidences. More principled conformal prediction based approaches could be adopted to calibrate LLM confidence for answers generated and use the measure for penalizing/promoting documents. We also defer this for future work. 
\section{Ethical Considerations and Risks}

We primarily propose a novel uncertainty based neighborhood aware retrieval approach for complex QA tasks. We use only publicly available QA datasets from Wikipedia that do not contain private or harmful information. While we employ LLMs for QA systems in our experiments, we do not prompt them in a format that would elicit harmful or biased information. All our prompts are reported in Appendix \ref{app:prompts}. Our tool is intended to enhance answering engines to aid users in their complex questions.

\bibliography{references}
\bibliographystyle{acl_natbib}

\clearpage
\appendix

\section{Dataset}
\label{sec:datasets}
We evaluate on MusiqueQA and 2WikimultihopQA released under CC-BY-4.0 LICENSE

    \textbf{2WikiMultihopQA (WQA)}: Wikipedia based multi-hop questions that consists of compositional queries and comparison based queries. We evaluate on 1200 compositional questions as done in self-ask \cite{self_ask} for a fair comparison. Example Question: Who was born later, Gideon Johnson or Holm Jølsen?

      \textbf{MuSiQue (MQA)} Comprises hard multi-hop compositional questions constructed from Wikipedia-based QA datasets. The dataset was constructed from multiple single-hop datasets to result in complex questions that mandate multistep reasoning.  MQA is a challenging dataset created to avoid possible shortcuts in existing multi-hop datasets like HotPotQA, where models do not perform compositional reasoning. We evaluate on 1252 compositional questions as done in self-ask \cite{self_ask} for a fair comparison.
    Example Question: What did the actress in My Fair Lady win a Tony for ?

\section{Baselines}
\label{app:baselines}
\textbf{PRF (w/o query understanding)}: We follow ANCE-PRF \cite{ance_prf}, where query representations are updated using all documents from first-stage retrieval. This makes a faulty assumption that top documents in first-stage retrieval are relevant. We make this much stronger by identifying golden documents directly from SPALDEv2 results and concatenate them to query. This updated query is encoded using ColBERTv2 to update query representations and employed to re-rank results from first-stage results. Note that this is an ideal scenario where gold documents are looked up to formulate a strong baseline to compare to \name{}.

\textbf{PRF (w query understanding)}: It is similar to the above setup further augmented with query understanding where the retrieval happens for each sub-question obtained from original question and the query representation is recomputed using gold documents from this combined retrieval pool across sub-questions. Using the new query representation, the combined retrieval pool is re-ranked using \textit{nreimers/mmarco-mMiniLMv2-L12-H384-v1} for fair comparison with \name{}. 
\section{Comparison of Retrievers}
\label{app:retrieval_perf}
\begin{table}[h!]
\small
\centering
\setlength{\tabcolsep}{3.0pt}
\begin{tabular}{p{2.8cm}cccc}
\toprule
\multirow{2}{*}{\textbf{Retriever}}  & \multicolumn{2}{c}{\textbf{MQA}} & \multicolumn{2}{c}{\textbf{WQA}}  \\
\cmidrule{2-5}
 & nDCG & R@10 &nDCG & R@10 \\
\midrule
\multirow{1}{*}{SPLADEV2} &  0.155 & 0.062  & 0.251   & 0.186   \\
PRF & 0.183 & 0.065& 0.271& 0.190 \\
$ \name{}_{R}$ & 0.179& \textbf{0.073}& \textbf{0.297} & \textbf{0.216} \\

\bottomrule
\end{tabular}
\caption{Retrieval performances on MQA and WQA  w/o query understanding. }
\label{table:comparison_retrieval}
\end{table}

\begin{table}[h!]
\small
\centering
\begin{tabular}{p{1.2cm}cccc}
\toprule
\multirow{2}{*}{\textbf{Retriever}}  & \multicolumn{2}{c}{\textbf{MQA}} & \multicolumn{2}{c}{\textbf{WQA}}  \\
\cmidrule{2-5}
 & nDCG & R@100 &nDCG@10 & R@100 \\
\midrule
\multirow{1}{*}{SPLADeV2} &  \textbf{0.155} & \textbf{0.181} & \textbf{0.251}   & \textbf{0.323}   \\
\midrule
\multirow{1}{*}{DPR} & 0.109 & 0.107  & 0.126 & 0.179  \\
\multirow{1}{*}{ANCE} & 0.140 & 0.115 & 0.212 & 0.223   \\

\midrule
\multirow{1}{*}{Tas-b} & 0.176 & 0.172 & 0.277 & 0.303 \\
\multirow{1}{*}{MPNet} & 0.163  & 0.149 & 0.222 & 0.253  \\
\multirow{1}{*}{Contriever} &  0.155 & 0.169 & 0.216 & 0.294   \\
\bottomrule
\end{tabular}
\caption{Performance of different retrieval models on MQA and WQA when documents are retrieved using original question.}
\label{table:retrieval_performance}
\end{table}

We evaluate different retrieval approaches for the original questions in WQA and MQA and report nDCG@10 and Recall@100 as shown in Table \ref{table:retrieval_performance}. We also compare $\name{}_R$ with PRF in w/o query understanding setup as observed in Table \ref{table:comparison_retrieval}. We observe that surprisingly sparse retrievers like SPLADEv2 provide superior performance as measured by nDCg@10 and Recall@100. This indicates that SPLADE is able to capture more relevant documents as measured by recall, and is also able to rank them higher as measured by nDCG compared to other retrievers. Hence, we employ SPLADEv2 as first-stage retriever for our approach and all baselines in this work for a fair comparison. From Table \ref{table:comparison_retrieval} we also observe that vanilla version of \name{} w/o query understanding still outperforms PRF based approaches which are limited due to possibility of query drift.
\begin{table}[!t]
    \centering
    \small
    \setlength{\tabcolsep}{2.0pt}
    \begin{tabular}{lccc}
    \toprule
     \textbf{Method}& \multicolumn{1}{c}{MQA}& \multicolumn{1}{c}{\wqa{}}\\

     \midrule

              \colorg \textbf{Methods (w/ query understanding)} & \colorg& \colorg& \colorg\\

     ReAct \cite{yao2023react} & 19.65 & 42.75  \\


    DecomP \cite{khot2023decomposed} & 22.66 & 44.62 \\

        SearChain \cite{xu2024searchinthechaininteractivelyenhancinglarge} & 25.71& 45.16  \\

            \self{} \cite{self_ask} & 24.84 & 45.83 \\



        \hline
                 \colorg \textbf{NAR (w/ query understanding) (ours)}& \colorg& \colorg& \colorg\\

        $\name{}_{R}$ & 28.11 & 47.67\\  
        \name{} & \textbf{32.75} $\dagger$& \textbf{54.83}$\dagger$ \\

     \bottomrule
    \end{tabular}
        \caption{Results across datasets where baselines employ re-ranking for each sub-question after first-stage retrieval. The model used for \name{} and other approaches is gpt-3.5-turbo unless otherwise specified.$\dagger$ indicates statistical significance over \self{} at 0.1 level.}
    \label{tab:ranking}
\end{table}

\section{Comparison with baselines with Re-Ranking }
\label{app:ranking}
The existing state-of-the-art query understanding approaches decompose a complex question and interactively query a retriever to get relevant evidence, followed by LLM based reasoning. We further enhance these baselines by re-ranking the contexts/evidences retrieved for each sub-question. We use \textit{nreimers/mmarco-mMiniLMv2-L12-H384-v1} as the ranker same as the scorer used in NAR for \name{} to ensure a fair comparison. We observe, from comparing the results in Table \ref{tab:ranking} and Table \ref{tab:main_result}, that re-ranking provided slight performance gains but not significant enough due to the bounded-recall problem where re-ranking is still limited by recall of documents retrieved in first stage using SPLADEv2. This further strengthens our hypothesis that NAR ($\name{}_R$ and LLM guided NAR (\name{}) help bridge the recall gap and lead to significant performance improvements for complex QA.

\section{Variation of number of documents}
\label{app:top_k}
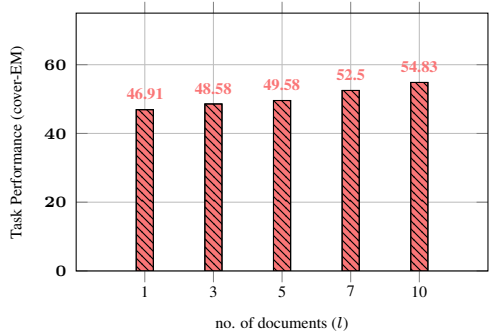
\begin{figure}[!t]
    \begin{subfigure}{.8\linewidth}
    \centering
    \begin{tikzpicture}
\edef\mylst{"46.91","48.58","49.58","52.5","54.83"}

    \begin{axis}[
            ybar=5pt,
            width=7cm,
            height=5cm,
            bar width=0.25,
            every axis plot/.append style={fill},
            grid=major,
            xtick={1, 2, 3, 4 , 5},
            xticklabels={ 1, 3, 5,7,10},
            xlabel={no. of documents ($l$)},
            ylabel style = {font=\tiny},
            xlabel style = {font=\tiny},
        yticklabel style = {font=\boldmath \tiny,xshift=0.05ex},
        xticklabel style ={font=\tiny,yshift=0.5ex},
            ylabel={Task Performance (cover-EM)},
            enlarge x limits=0.25,
            ymin=0,
            ymax=75,
            nodes near coords style={font=\tiny,align=center,text width=2em},
            legend cell align={left},
            legend pos=north west,
            legend columns=-1,
    extra y tick style={grid=major,major grid style={thick,draw=black}}, 
            legend style={/tikz/every even column/.append style={column sep=0.5cm}},
        ]
        \addplot+[
            ybar,
            plotColor1*,
            draw=black,
            nodes near coords=\pgfmathsetmacro{\mystring}{{\mylst}[\coordindex]}\textbf{\mystring},
    nodes near coords align={vertical},
            postaction={
                    pattern=north west lines
                },
        ] plot coordinates {
                (1,46.91)
                (2,48.58)
                (3,49.58)
                (4,52.5)
                (5,54.83)

            };

    \end{axis}

\end{tikzpicture}
    \end{subfigure}
\caption{QA performance for different values of l in top-$l$ documents used for reasoning where $l$=\{1,3,5,7,10}\}. $l$=10 works best.
\label{fig:top_k}
\end{figure}

We vary the number of top-$l$ documents used for downstream reasoning with LLM for \name{}, where $l = \{ 1,3,5,7,10\}$ for WQA as shown in Figure \ref{fig:top_k}. We observe that $l=10$ yields the best performance. We also observe that performance variation across different values of l is not huge for \name{} demonstrating that \name{} captures relevant documents in top-10 results compared to other approaches. We also use top-10 documents for LLM reasoning for all other baselines for a fair comparison.
 
\begin{table}[hbt!]

\begin{tcolorbox}[title= Meta-Reasoner Prompt]
\small
\textbf{Instruction}:\texttt{Follow the given examples and Given the question and context, reasoning path, think step by step extract key segments from given evidence relevant to question and give rationale, by forming your own reasoning path preceded by [Answer]: and output final answer for the question using information from given evidences and give concise precise answer preceded by [Final Answer]:.}

\paragraph{\textbf{Exemplars}}: \{\} \\

Given the above examples and 
\\\textbf{Existing reasoner path}: \{Reasoning path from $\name{}_R$\}, \\
\textbf{the evidence} : \{top-$l$ evidences across sub-questions\} \\ and use the most relevant information for the question from the most relevant evidence from given Evidence: and form your own correct reasoning path to derive the answer thinking step by step preceded by [Answer]: and subsequently give final answer as shown in above examples preceded by [Final Answer]: for \\ \textbf{the Question}: \{Test Question\}







\end{tcolorbox}
\captionof{figure}{Meta-reasoner prompt}
\label{fig:meta} 
\end{table}

\section{Prompts}
\label{app:prompts}
\begin{table*}[hbt!]

\begin{tcolorbox}[title= \self{} MusiqueQA Prompt]
\small
\textbf{Instruction}:\texttt{Follow the given examples and Given the question determine if followup questions are needed and decompose the original question.}

\paragraph{\textbf{Exemplars}}:







Question: When does monsoon season end in the state the area code 575 is located? \\
Are follow up questions needed here: Yes. \\
Follow up: Which state is the area code 575 located in?\\
Intermediate Answer: The area code 575 is located in New Mexico. \\
Follow up: When does monsoon season end in New Mexico? \\
Intermediate Answer: Monsoon season in New Mexico typically ends in mid-September. 

[Final Answer]: mid-September. 
\\

Question: What is the current official currency in the country where Ineabelle Diaz is a citizen? \\
Are follow up questions needed here: Yes. \\
Follow up: Which country is Ineabelle Diaz a citizen of? \\
Intermediate Answer: Ineabelle Diaz is from Peurto Rico, which is in the United States of America. \\
Follow up: What is the current official currency in the United States of America? \\
Intermediate Answer: The current official currency in the United States is the United States dollar.
[Final Answer]: United States dollar.
\\

Question: Where was the person who founded the American Institute of Public Opinion in 1935 born?
Are follow up questions needed here: Yes.
Follow up: Who founded the American Institute of Public Opinion in 1935?
Intermediate Answer: George Gallup.
Follow up: Where was George Gallup born?
Intermediate Answer: George Gallup was born in Jefferson, Iowa.

[Final Answer]: Jefferson.
\\

Question: What is the sports team the person played for who scored the first touchdown in Superbowl 1?  \\
Are follow up questions needed here: Yes. \\
Follow up: Which player scored the first touchdown in Superbowl 1? \\
Intermediate Answer: Max McGee. \\
Follow up: Which sports team did Max McGee play for? \\
Intermediate Answer: Max McGee played for the Green Bay Packers.

[Final Answer]: Green Bay Packers.
\\

Question: The birth country of Jayantha Ketagoda left the British Empire when? \\
Are follow up questions needed here: Yes. \\
Follow up: What is the birth country of Jayantha Ketagoda? \\
Intermediate Answer: Sri Lanka. \\
Follow up: When did Sri Lanka leave the British Empire? \\
Intermediate Answer: Sri Lanka left the British Empire on February 4, 1948. 

[Final Answer]: February 4, 1948 \\
                      
\paragraph{\textbf{Input}}: Based on above examples, given the Question: {} determine, Are followup questions needed here ?

\end{tcolorbox}
\captionof{figure}{Example of In-context learning for MusiqueQA (MQA) through \self{} based prompting of LLMs}
\label{prompt:mqa} 
\end{table*}

\begin{table*}[hbt!]

\begin{tcolorbox}[title= \self{} 2WikiMultiHopQA Prompt]
\small
\textbf{Instruction}:\texttt{Follow the given examples and Given the question determine if followup questions are needed and decompose the original question.}

\paragraph{\textbf{Exemplars}}:







Question: Who lived longer, Theodor Haecker or Harry Vaughan Watkins? \\
                Are follow up questions needed here: Yes.\\
Follow up: How old was Theodor Haecker when he died?\\
Intermediate Answer: Theodor Haecker was 65 years old when he died.\\
Follow up: How old was Harry Vaughan Watkins when he died?\\
Intermediate Answer: Harry Vaughan Watkins was 69 years old when he died.

[Final Answer]: Harry Vaughan Watkins.
\\
Question: Why did the founder of Versus die? \\
Are follow up questions needed here: Yes.\\
Follow up: Who founded Versus?\\
Intermediate Answer: Gianni Versace.\\
Follow up: Why did Gianni Versace die?\\
Intermediate Answer: Gianni Versace was shot and killed on the steps of his Miami Beach mansion on July 15, 1997.
[Final Answer]: Shot.
\\
Question: Who is the grandchild of Dambar Shah? \\
Are follow up questions needed here: Yes. \\
Follow up: Who is the child of Dambar Shah? \\
Intermediate Answer: Dambar Shah (? - 1645) was the king of the Gorkha Kingdom. He was the father of Krishna Shah. \\
Follow up: Who is the child of Krishna Shah? \\
Intermediate Answer: Krishna Shah (? - 1661) was the king of the Gorkha Kingdom. He was the father of Rudra Shah. 
[Final Answer]: Rudra Shah.
\\
Question: Are both director of film FAQ: Frequently Asked Questions and director of film The Big Money from the same
country? \\
Are follow up questions needed here: Yes. \\
Follow up: Who directed the film FAQ: Frequently Asked Questions?\\
Intermediate Answer: Carlos Atanes.\\
Follow up: Who directed the film The Big Money?
Intermediate Answer: John Paddy Carstairs.
Follow up: What is the nationality of Carlos Atanes?
Intermediate Answer: Carlos Atanes is Spanish.
Follow up: What is the nationality of John Paddy Carstairs?\\
Intermediate Answer: John Paddy Carstairs is British.
[Final Answer]: No.
\\
Question: Who was the maternal grandfather of George Washington? \\
Are follow up questions needed here: Yes.
Follow up: Who was the mother of George Washington?\\
Intermediate answer: The mother of George Washington was Mary Ball Washington. \\
Follow up: Who was the father of Mary Ball Washington? \\
Intermediate answer: The father of Mary Ball Washington was Joseph Ball.
[Final Answer]: Joseph Ball 
\\

\paragraph{\textbf{Input}}: Based on above examples, given the Question: {} determine, Are followup questions needed here ?

\end{tcolorbox}
\captionof{figure}{Example of In-context learning for 2WikiMultiHopQA through \self{} based prompting of LLMs }
\label{prompt:wqa} 
\end{table*}

\begin{table*}[hbt!]

\begin{tcolorbox}[title= SearChain Prompt]
\small
\textbf{Instruction}:\texttt{Solve a question answering task with interleaving Thought, Action, Observation steps. Thought can reason about the current situation, and Action can be three types: 
(1) Search[entity], which searches the exact entity on Wikipedia and returns the first paragraph if it exists. If not, it will return some similar entities to search. \\
(2) Lookup[keyword], which returns the next sentence containing keyword in the current passage. \\
(3) Finish[answer], which returns the answer and finishes the task.}

\paragraph{\textbf{Exemplars}}:







Question: When does monsoon season end in the state the area code 575 is located? \\
Are follow up questions needed here: Yes. \\
Follow up: Which state is the area code 575 located in?\\
Intermediate Answer: The area code 575 is located in New Mexico. \\
Follow up: When does monsoon season end in New Mexico? \\
Intermediate Answer: Monsoon season in New Mexico typically ends in mid-September. 

[Final Answer]: mid-September. 
\\

Construct a global reasoning chain for this complex [Question] : " \{\} " You should generate a query to the search engine based on
        what you already know at each step of the reasoning chain, starting with [Query]. \\
                        If you know the answer for [Query], generate it starting with [Answer]. \\
                        You can try to generate the final answer for the [Question] by referring to the [Query]-[Answer] pairs, starting with [Final
                        Answer].\\
                        If you don't know the answer, generate a query to search engine based on what you already know and do not know, starting with
                        [Unsolved Query]. \\
                        For example:
                        [Question]: "Where do greyhound buses that are in the birthplace of Spirit If...'s performer leave from? "

                        [Query 1]: Who is the performer of Spirit If... ? \\
                        If you don't know the answer:

                        [Unsolved Query]: Who is the performer of Spirit If... ? \\
                        If you know the answer:
                        [Answer 1]: The performer of Spirit If... is Kevin Drew.
                        [Query 2]: Where was Kevin Drew born? \\
                        If you don't know the answer:
                        [Unsolved Query]: Where was Kevin Drew born? \\
                        If you know the answer:
                        [Answer 2]: Toronto. 
                        
                        [Query 3]: Where do greyhound buses in Toronto leave from? \\
                        If you don't know the answer:
                        [Unsolved Query]: Where do greyhound buses in Toronto leave from? \\
                        If you know the answer:
                        [Answer 3]: Toronto Coach Terminal. 
                        
                        [Final Content]: The performer of Spirit If... is Kevin Drew [1]. Kevin Drew was born in Toronto [2]. Greyhound buses in
                        Toronto leave from Toronto
                        Coach Terminal [3]. So the final answer is Toronto Coach Terminal. 
                        
                        [Final Answer]: Toronto Coach Terminal

                        [Question]:"Which magazine was started first Arthur’s Magazine or First for Women?"

                        [Query 1]: When was Arthur’s Magazine started?

                        [Answer 1]: 1844.

                        [Query 2]: When was First for Women started?
                        [Answer 2]: 1989

                        [Final Content]: Arthur’s Magazine started in 1844 [1]. First for Women started in 1989 [2]. So Arthur’s Magazine was started
                        first. So the answer is Arthur’s Magazi

                        [Final Answer]: Arthur’s Magazi.
                      [\dots]

\end{tcolorbox}
\captionof{figure}{Searchain prompt used for complex QA \cite{xu2024searchinthechaininteractivelyenhancinglarge}}
\label{prompt:searchain} 
\end{table*}

\begin{table*}[hbt!]

\begin{tcolorbox}[title= DecomP Prompt]
\small
\textbf{Instruction}:\texttt{Solve a question answering task with interleaving retrieval 
Follow the given examples and Given the question and context, use relevant evidence for each subquestion from given list of evidences, answer the subquestion and output final answer for the question using information in the context, answers to subquestions and give final answer strictly preceded by  [Final Answer]:}

\paragraph{\textbf{Exemplars}}:







[Question]: In which country did this Australian who was detained in Guantanamo Bay detention camp
and published Guantanamo: My Journey receive para-military training?

Subquestion 1: (select) [retrieve odqa] Who is the Australian who was detained in Guantanamo Bay detention
camp and published ”Guantanamo: My Journey”?

Answer: {”titles”: [”Guantanamo: My Journey”, ”Bismullah v. Gates”, ”Guantanamo Bay detention
camp”], ”answer”: [”David Hicks”]}

Subquestion 1: (select) [retrieve odqa] In which country did David Hicks receive his para-military training?

Answer: {”titles”: [”John Adams Project”, ”Camp Echo (Guantanamo Bay)”, ”Guantanamo Bay
Museum of Art and History”, ”David Hicks”], ”answer”: [”Afghanistan”]}

Subquestion 3: (select) [multihop titleqa] Titles: [”Guantanamo: My Journey”, ”Bismullah v. Gates”, ”
Guantanamo Bay detention camp”, ”John Adams Project”, ”Camp Echo (Guantanamo Bay)”,
”Guantanamo Bay Museum of Art and History”, ”David Hicks”]. 

Question: In which country
did this Australian who was detained in Guantanamo Bay detention camp and published ``
Guantanamo My Journey" receive para-military training?

[Final Answer]: Afghanistan
\\

[Question]: who is older Jeremy Horn or Renato Sobral ?
Subquestion 1: (select) [retrieve odqa] When was Jeremy Horn born?

Answer: {”titles”: [”Zaza Tkeshelashvili”, ”Jeremy Horn”, ”Jeremy Horn (singer)”, ”Ricardo Arona”], ”
answer”: [”August 25, 1975”]}

Subquestion 2: (select) [retrieve odqa] When was Renato Sobral born?

Answer: {”titles”: [”Brian Warren”, ”Renato Sobral”], ”answer”: [”September 7, 1975”]}

Subquestion 3: (select) [multihop titleqa] Titles: [”Zaza Tkeshelashvili”, ”Jeremy Horn”, ”Jeremy Horn (
singer)”, ”Ricardo Arona”, ”Brian Warren”, ”Renato Sobral”]. Question: who is older Jeremy
Horn or Renato Sobral ?

[Final Answer]: Jeremy Horn
\\

[Question]: What was the 2014 population of the city where Lake Wales Medical Center is located?

Subquestion 1: (select) [retrieve odqa] Lake Wales Medical Center is located in what city?

Answer: {”titles”: [”Baylor College of Medicine”, ”Lake Wales Medical Center”, ”Tufts University
School of Medicine”, ”Hanford Community Medical Center”], ”answer”: [”Polk County,
Florida”]}

Subquestion 2: (select) [retrieve odqa] What was the population of Polk County in 2014?

Answer: {”titles”: [”Banner University Medical Center Tucson”, ”Lake Wales, Florida”], ”answer”:
[”15,140”]}

Subquestion 3: (select) [multihop titleqa] Titles: [”Baylor College of Medicine”, ”Lake Wales Medical Center
”, ”Tufts University School of Medicine”, ”Hanford Community Medical Center”, ”Banner
University Medical Center Tucson”, ”Lake Wales, Florida”]. Question: What was the 2014
population of the city where Lake Wales Medical Center is located?

[Final Answer]: 15,140
\\

[Question]: Nobody Loves You was written by John Lennon and released on what album that was issued
by Apple Records, and was written, recorded, and released during his 18 month separation
from Yoko Ono?

Subquestion 1: (select) [retrieve odqa] What album was issued by Apple Records, and written, recorded, and
released during John Lennon’s 18 month separation from Yoko Ono?

Answer: {”titles”: [”John Lennon/Plastic Ono Band”, ”Milk and Honey (album)”, ”Walls and Bridges”],
”answer”: [”Walls and Bridges”]}

Subquestion 2: (select) [retrieve odqa] Nobody Loves You was written by John Lennon on what album?

Answer: {”titles”: [”John Lennon Museum”, ”Nobody Loves You (When You’re Down and Out)”], ”
answer”: [”Walls and Bridges”]}

Subquestion 3: (select) [multihop titleqa] Titles: [”John Lennon/Plastic Ono Band”, ”Milk and Honey (album)

[Final Answer]: Walls and Bridges
                      [\dots]

\paragraph{\textbf{Input}:} "Follow the above examples, and use the given Evidence: "+\{\}+"  and use the information from the evidence to answer the original question  and give final answer strictly preceded by [Final Answer]: for the   Question: \{\}

\end{tcolorbox}
\captionof{figure}{DeComP prompt \cite{khot2023decomposed}}
\label{prompt:decomp} 
\end{table*}

\begin{table*}[hbt!]

\begin{tcolorbox}[title= ReAct Prompt]
\small
\textbf{Instruction}:\texttt{Solve a question answering task with interleaving Thought, Action, Observation steps. Thought can reason about the current situation, and Action can be three types: 
(1) Search[entity], which searches the exact entity on Wikipedia and returns the first paragraph if it exists. If not, it will return some similar entities to search.
(2) Lookup[keyword], which returns the next sentence containing keyword in the current passage.
(3) Finish[answer], which returns the answer and finishes the task.:}

\paragraph{\textbf{Exemplars}}:







Question: What is the elevation range for the area that the eastern sector of the Colorado orogeny extends into?\\
                Thought 1: I need to search Colorado orogeny, find the area that the eastern sector of the Colorado orogeny extends into, then find the elevation range of the area.\\
                Action 1: Search[Colorado orogeny]\\
                Observation 1: The Colorado orogeny was an episode of mountain building (an orogeny) in Colorado and surrounding areas.\\
                Thought 2: It does not mention the eastern sector. So I need to look up eastern sector.\\
                Action 2: Lookup[eastern sector]\\
                Observation 2: (Result 1 / 1) The eastern sector extends into the High Plains and is called the Central Plains orogeny.\\
                Thought 3: The eastern sector of Colorado orogeny extends into the High Plains. So I need to search High Plains and find its elevation range.\\
                Action 3: Search[High Plains]\\
                Observation 3: High Plains refers to one of two distinct land regions:\\
                Thought 4: I need to instead search High Plains (United States).\\
                Action 4: Search[High Plains (United States)]\\
                Observation 4: The High Plains are a subregion of the Great Plains. From east to west, the High Plains rise in elevation from around 1,800 to 7,000 ft (550 to 2,130 m).[3]\\
                Thought 5: High Plains rise in elevation from around 1,800 to 7,000 ft, so the answer is 1,800 to 7,000 ft.\\
                Action 5: Finish[1,800 to 7,000 ft]

                [Final Answer]: 1,800 to 7,000 ft \\
                Question: Musician and satirist Allie Goertz wrote a song about the \"The Simpsons\" character Milhouse, who Matt Groening named after who?\\
                Thought 1: The question simplifies to \"The Simpsons\" character Milhouse is named after who. I only need to search Milhouse and find who it is named after.\\
                Action 1: Search[Milhouse]\\
                Observation 1: Milhouse Mussolini Van Houten is a recurring character in the Fox animated television series The Simpsons voiced by Pamela Hayden and created by Matt Groening.\\
                Thought 2: The paragraph does not tell who Milhouse is named after, maybe I can look up \"named after\".\\
                Action 2: Lookup[named after]\\Observation 2: (Result 1 / 1) Milhouse was named after U.S. president Richard Nixon, whose middle name was Milhous. \\
                Thought 3: Milhouse was named after U.S. president Richard Nixon, so the answer is Richard Nixon.\\
                Action 3: Finish[Richard Nixon]\\Question: Which documentary is about Finnish rock groups, Adam Clayton Powell or The Saimaa Gesture?\\
                Thought 1: I need to search Adam Clayton Powell and The Saimaa Gesture, and find which documentary is about Finnish rock groups.\\
                Action 1: Search[Adam Clayton Powell]\\
                Observation 1: Could not find [Adam Clayton Powell]. Similar: ['Adam Clayton Powell III', 'Seventh Avenue (Manhattan)', 'Adam Clayton Powell Jr. State Office Building', 'Isabel Washington Powell', 'Adam Powell', 'Adam Clayton Powell (film)', 'Giancarlo Esposito'].\\
                Thought 2: To find the documentary, I can search Adam Clayton Powell (film).\\
                Action 2: Search[Adam Clayton Powell (film)]\\
                Observation 2: Adam Clayton Powell is a 1989 American documentary film directed by Richard Kilberg.\\
                The film is about the rise and fall of influential African-American politician Adam Clayton Powell Jr.[3][4] It was later aired as part of the PBS series The American Experience.\\
                Thought 3: Adam Clayton Powell (film) is a documentary about an African-American politician, not Finnish rock groups. So the documentary about Finnish rock groups must instead be The Saimaa Gesture.\\
                Action 3: Finish[The Saimaa Gesture]

                [Final Answer]: The Saimaa Gesture \\

[\dots]

\paragraph{\textbf{Input}:} "Follow the above examples, and use the given Evidence: "+\{\}+" \\ and use the information from the evidence to answer the original question  and give final answer strictly preceded by [Final Answer]: for the   Question: \{\}
\end{tcolorbox}
\captionof{figure}{ReAct prompt \cite{khot2023decomposed}}
\label{prompt:react} 
\end{table*}

We present the prompts used for the baselines and \name{} for the query understanding, which aids in reasoning. These prompts are used for all LLMs. The prompt employed for \self{} along with demonstration samples/exemplars are as shown in Figure \ref{prompt:wqa} (WQA) and Figure \ref{prompt:mqa} (MQA), meta-reasoner prompt template (Figure \ref{fig:meta}), SearChain (Figure \ref{prompt:searchain}) and DecomP (Figure \ref{prompt:decomp} and ReAct \ref{prompt:react}. We use a prompt similar to \self{} for \name{} main results in Table \ref{tab:main_result}, but we also demonstrate the generality of our approach by applying query understanding approaches ReAct and SearChain in Table \ref{tab:reasoning_substrates}.

\end{document}